%% file: Bishop-DrivenJC.tex
\documentclass[aps,prl,reprint,USenglish,floatfix,showpacs]{revtex4-1}
\usepackage{ae}
\usepackage{ifdraft}
\usepackage{babel}
\usepackage[rgb]{xcolor}
\usepackage[final]{graphicx}
\usepackage{amsmath,amssymb}

\input{macros.tex}

\usepackage[final]{hyperref}
\usepackage[all]{hypcap}
\begin{document}
\author{Lev S. Bishop}
\author{Eran Ginossar}
\author{S. M. Girvin}
\affiliation{Physics Department, Yale University, New Haven, Connecticut 06520, USA}
\title{Response of the Strongly-Driven Jaynes--Cummings Oscillator}
\begin{abstract}
We analyze the Jaynes-Cummings model of quantum optics, in the strong-dispersive regime. In the bad cavity limit and on timescales short compared to the atomic coherence time, the dynamics are those of a nonlinear oscillator. A steady-state non-perturbative semiclassical analysis exhibits a finite region of bistability delimited by a pair of critical points, unlike the usual dispersive bistability from a Kerr nonlinearity. This analysis explains our quantum trajectory simulations that show qualitative agreement with recent experiments from the field of circuit quantum electrodynamics.
\end{abstract}
\date{May 3, 2010}
\maketitle

The Jaynes--Cummings~(JC) Hamiltonian provides a quantum model for a two-level system (qubit) interacting with a quantized electromagnetic mode. It is widely applicable to experiments with natural atoms~\cite{haroche_raimond_exploring,guerlin_progressive_2007,gleyzes_quantum_2007,boca_observation_2004,brennecke_cavity_2007, maunz_normal-mode_2005,gupta_cavity_2007,sauer_cavity_2004} as well as for solid-state `artificial atoms'~\cite{reithmaier_strong_2004,yoshie_vacuum_2004,wallra_strong_2004}.
The JC Hamiltonian can be diagonalized analytically, but in the presence of a drive $\xi(t)$ and dissipation the open-system model becomes non-trivial, with the effective behavior depending strongly on the specific parameter regime. The case where the cavity relaxation rate $\kappa$ greatly exceeds the two-level dissipation and dephasing rates $\gamma$, $\gamma_\phi$ is known as the bad cavity limit. The \emph{strong dispersive} regime~\cite{gambetta_qubit-photon_2006,schuster_resolving_2007} of the JC model describes the situation that the presence of the qubit causes the cavity frequency to be shifted by an amount $\chi$ much greater than the cavity linewidth. Recent experiments~\cite{reedout_expt}, which operate in both the strong dispersive regime and the bad cavity limit, show a nontrivial response under conditions of strong drive, arousing interest due to its usefulness for high-fidelity qubit readout. A characteristic feature of the JC model is that for very high excitation number $N\gg 1$, the excitation number-dependent shift obeys $\chi(N) \rightarrow 0$: the transition frequency returns to the bare cavity frequency. In the presence of dissipation this happens effectively when $\chi(N) \lesssim \kappa$, and for all larger $N$ the response of the system is linear with respect to the drive. We describe this behavior as setting in at an excitation number $\Nbare$, with the definition $\chi(\Nbare)=\kappa$. In the strong dispersive regime we have $\Nbare\gg\Ncrit$, where $\Ncrit$ as usual denotes the excitation level where the dispersive approximation breaks down (defined below). The latter inequality has an important consequence for the theory: a perturbative expansion in the small parameter $N/\Ncrit$, typically useful~\cite{boissonneault_nonlinear_2008,*boissonneault_dispersive_2009} in the dispersive regime, is not applicable for the interesting regime $N>\Nbare$ where the system regains the linear response.

In this paper we consider the JC model under very strong driving, such that $N\gg\Ncrit$. Our main result is that there exists a threshold drive $\xi_\text{C2}$ at which the photon occupation increases by several orders of magnitude over a small range of the drive amplitude. We perform both quantum trajectory simulations and a non-perturbative semiclassical analysis, including the drive and the cavity damping. Our results are in qualitative agreement with recent experiments~\cite{reedout_expt} for a circuit quantum electrodynamics~(QED) device~\cite{dicarlo_GHZ} containing 4 transmon~\cite{koch_charge-insensitive_2007,schreier_suppressing_2008} qubits, demonstrating that the JC model captures the essential physics despite making an enormous simplification of the full system Hamiltonian.

The behavior of the JC nonlinearity goes beyond the Kerr nonlinearity that is often considered. Dispersive bistability~\cite{marburger_theory_1978} from a Kerr nonlinearity is well-known in atomic cavity QED~\cite{gibbs_differential_1976}. It has been implemented in the solid state via the nonlinearity of a Josephson junction~\cite{siddiqi_rf-driven_2004}, and in the circuit QED architecture has produced high-fidelity readout of qubits~\cite{siddiqi_dispersive_2006,*boulant_quantum_2007,*metcalfe_measuringdecoherence_2007,mallet_single-shot_2009}. Similar schemes have been implemented with nonlinear micromechanical resonators~\cite{almog_high_2006}. However, unlike the Kerr anharmonicity, the JC anharmonicity does not remain constant but rather diminishes toward zero as the cavity occupation is increased. As a result, for sufficiently strong drive the response of the JC model must return to the linear response of the bare cavity. Instead of coherent driving, an alternative way to saturate the qubit and cause the JC system response to return to the bare cavity response is to couple the system to a bath at elevated temperature, as has been investigated theoretically~\cite{rau_cavity_2004} and experimentally~\cite{fink_quantum-to-classical_2010}.
\begin{figure}
\includegraphics{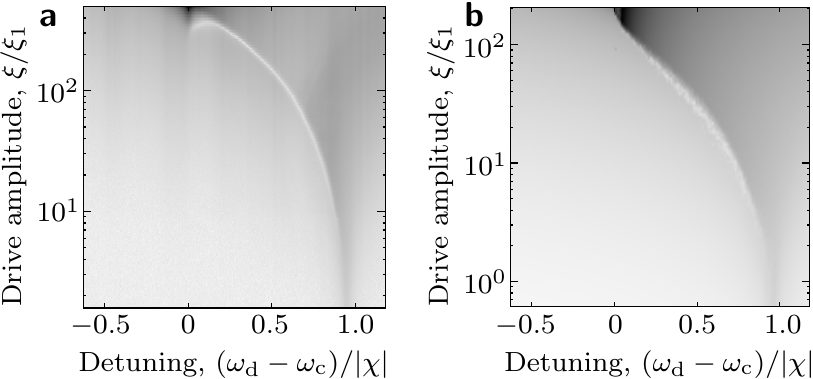}
\caption{Transmitted heterodyne amplitude $\lvert\langle a\rangle\rvert$ as a function of drive detuning (normalized by the dispersive shift $\chi=g^2/\delta$) and drive amplitude (normalized by the amplitude to put $n=1$ photon in the cavity in linear response, $\xi_1=\kappa/\sqrt{2}$). Dark colors indicate larger amplitudes.
(a)~Experimental data~\cite{matt-pc}, for a sample with a cavity at $9.07\,\text{GHz}$ and 4 transmon qubits at $7.0, 7.5, 8.0, 12.3\,\text{GHz}$. All qubits are initialized in their ground state, and the signal is integrated for the first $400\,\text{ns}\simeq4/\kappa$ after switching on the drive.
(b)~Numerical results for the JC model of Eq.~\eqref{eq:master}, with qubit fixed to the ground state and effective parameters $\delta/2\pi=-1.0\,\text{GHz}$, $g/2\pi=0.2\,\text{GHz}$, $\kappa/2\pi=0.001\,\text{GHz}$. These parameters are only intended as representative numbers for circuit QED and were not optimized against the data of panel~(a). Hilbert space is truncated at 10,000 excitations (some truncation artifacts are visible for the strongest drive), and results are shown for time $t=2.5/\kappa$.\label{fig:latch000}}
\end{figure}

Our analysis is applicable to any experiment that can reach the strong dispersive limit and drive sufficiently strongly, but for concreteness we adopt the language and typical parameters of the field of circuit QED\@. We write the driven JC Hamiltonian
\be\begin{split}
  H&=\omc a^{\dag}a+\frac{\omq}{2} \sigma_z + g(a \sigma_+ + a^\dag \sigma_-) \\
  &\quad+\frac{\xi}{\sqrt{2}}(a+a^\dag)\cos(\omd t),
\end{split}\ee
with cavity frequency $\omc/2\pi$, qubit frequency $\omq/2\pi$, coupling strength $g$, drive amplitude $\xi$ and drive frequency $\omd$. Operating in the strong-dispersive bad-cavity regime defines a hierarchy of scales
\be\label{eq:regime}
    \gamma,\gamma_\phi\ll\kappa\ll g^2/\delta\ll g\ll \delta\ll\omc ,
\ee
where $\delta=\omq-\omc$ is the qubit-cavity detuning.
We can make the standard transformation~\cite{canonicaldressing} to decouple the qubit and cavity
\be\label{eq:H_car}\begin{split}
  \tilde{H}&=T^{-1} H T= \omc a^\dag a+ (\omc-\Delta)\frac{\sigma_z}{2}\\
  &\quad+\frac{\xi}{\sqrt{2}}(a+a^{\dag})\cos(\omd t) ,
\end{split}\ee
dropping terms from the transformed drive that are suppressed as $\bigO(n^{-1/2})$ and $\bigO(g/\delta)$. The resulting Hamiltonian would be trivial were it not for the fact that the transformation $T$, defined by
\begin{align}
  T&=\exp[-\theta(4N)^{-1/2}(a \sigma_+ + a^\dag \sigma_-)] , \\
  \sin\theta&=-2g N^{1/2}/\Delta,\quad \cos\theta=\delta/\Delta , \\
  \Delta&=(\delta^2+4g^2 N)^{1/2} \label{eq:Delta},
\end{align}
depends on $N$, the total number of excitations:
\be
    N=a\dg a +\frac{\sigma_z}{2} + \frac{1}{2} .
\ee
For photon decay at rate $\kappa$ we can write the decoupled quantum master equation after dropping small terms,
\be\label{eq:master}
    \dot{\rho}=-\rmi[\tilde{H},\rho]+\kappa([a\rho,a\dg]+[a,\rho a\dg])/2,
\ee
which we integrate numerically in a truncated Hilbert space using the method of quantum trajectories, after making the rotating wave approximation~(RWA) with respect to the drive. The experiments we wish to describe are performed on a timescale short compared to the qubit decoherence times $\gamma^{-1}$, $\gamma_\phi^{-1}$ and we therefore treat $\sigma_z$ as a constant of motion. The remaining degree of freedom constitutes a Jaynes--Cummings oscillator. Note that the qubit relaxation and dephasing terms that we have dropped involve the $\sigma_\pm$ and $\sigma_z$ operators and would transform in a nontrivial way under the decoupling transformation $T$~\cite{boissonneault_nonlinear_2008,*boissonneault_dispersive_2009}. The results of the numerical integration for $\sigma_z=-1$ are compared with recent experimental data~\cite{matt-pc} in Fig.~\ref{fig:latch000}, where we show the average heterodyne amplitude $\lvert\langle a \rangle\rvert$ as a function of drive frequency and amplitude. Despite the presence of 4 qubits in the device, the fact that extensions beyond a two-level model would seem necessary since higher levels of the transmons are certainly occupied for such strong driving~\footnote{We performed auxiliary simulations showing that approximately 10 transmon levels will be required to simulate the experiment quantitatively.}, and despite the fact that the Rabi Hamiltonian might seem more appropriate for such large photon occupation, $\sqrt{N}g\sim\omc$, nevertheless the JC model qualitatively reproduces the features of the experiment~\footnote{We emphasize that the effective parameters in the simulation are of the same magnitude as in the experiments but we do not expect any quantitative correspondence.}. In particular, for weak driving we see a response as expected at the dispersively shifted cavity frequency $\omc-\chi$, with $\chi=g^2/\delta$, which shifts towards lower frequencies as the drive increases. For stronger driving a dip appears in the response, which we interpret as a consequence of plotting the ensemble-averaged $\lvert\langle a\rangle\rvert$ in the classically bistable region, as we discuss below. For increasing drive the dip shifts to lower frequencies; finally for the strongest driving, the response becomes centered at the bare cavity frequency $\omc/2\pi$ and is single-peaked and extremely strong. We note that both the experiment and numerical integration are terminated at a transient time of only a few cavity lifetimes, and we have checked that the numerical response is substantially different for the steady state.

Due to the large number of photons in the system, it is possible to form a semiclassical model, similar to Refs.~\cite{alsing_spontaneous_1991,*kilin_single-atom_1991,peano_dynamical_2010}. This will be a good approximation in the case that ratio of the anharmonicity of the dispersive Hamiltonian to the decay rate is such that the $N-1\leftrightarrow N$ photon peak overlaps well with the $N\leftrightarrow N+1$ photon peak, $N\gg\Nsc$, where $\Nsc=g^4/\kappa\delta^3$ (for the parameters of Fig.~\ref{fig:latch000}, $\Nsc=1.6$). In the opposite limit we will see photon blockade and associated effects, as in Ref.~\cite{bishop_nonlinear_2009}. Recently it was shown that it is possible to have a coexistence of both the semiclassical and quantum solutions for a certain range of parameters of the system and drive~\cite{blammo}. The semiclassical model will remain valid for $N>\Ncrit$, where a perturbative expansion of the Hamiltonian (\ref{eq:H_car}) in terms of $N/\Ncrit$ fails to converge, where $\Ncrit=\delta^2/4g$. We rewrite the Hamiltonian Eq.~\eqref{eq:H_car} using canonical variables $X=\sqrt{1/2}(a\dg+a)$ and $P=\rmi\sqrt{1/2}(a\dg-a)$,
\be\begin{split}
  \tilde{H}&=\frac{\omc}{2}(X^2+P^2+\sigma_z)+\xi X \cos(\omd t)\\
  &-\frac{\sigma_z}{2}\sqrt{2g^2(X^2+P^2+\sigma_z)+\delta^2}.
\end{split}\ee
The semiclassical approximation consists of treating $X$ and $P$ as numbers, and the effect of cavity relaxation is incorporated through a damping term proportional to $\kappa$. We solve for the steady state, treating $X^2+P^2$ as a constant (thus we ignore harmonic generation), giving a nonlinear equation for the amplitude $A=\sqrt{X^2+P^2}$
\be\label{eq:classic}
    A^2=\frac{\omc^2\xi^2}{\left[\omd^2-(\omc-\chi(A))^2\right]^2+\kappa^2\omd^2}
\ee
with amplitude-dependent frequency shift $\chi(A)$, given by
\be\label{eq:chi}
   \chi(A)=\sigma_z \frac{g^2}{\sqrt{2g^2(A^2+\sigma_z)+\delta^2}}.
\ee
This reproduces for small driving the usual dispersive shift $\chi(0)\simeq\pm g^2/\delta$ and for strong driving shows the saturation effect $\lim_{A\to\infty}\chi(A)=0$.
\begin{figure}
\includegraphics[width=0.8\columnwidth]{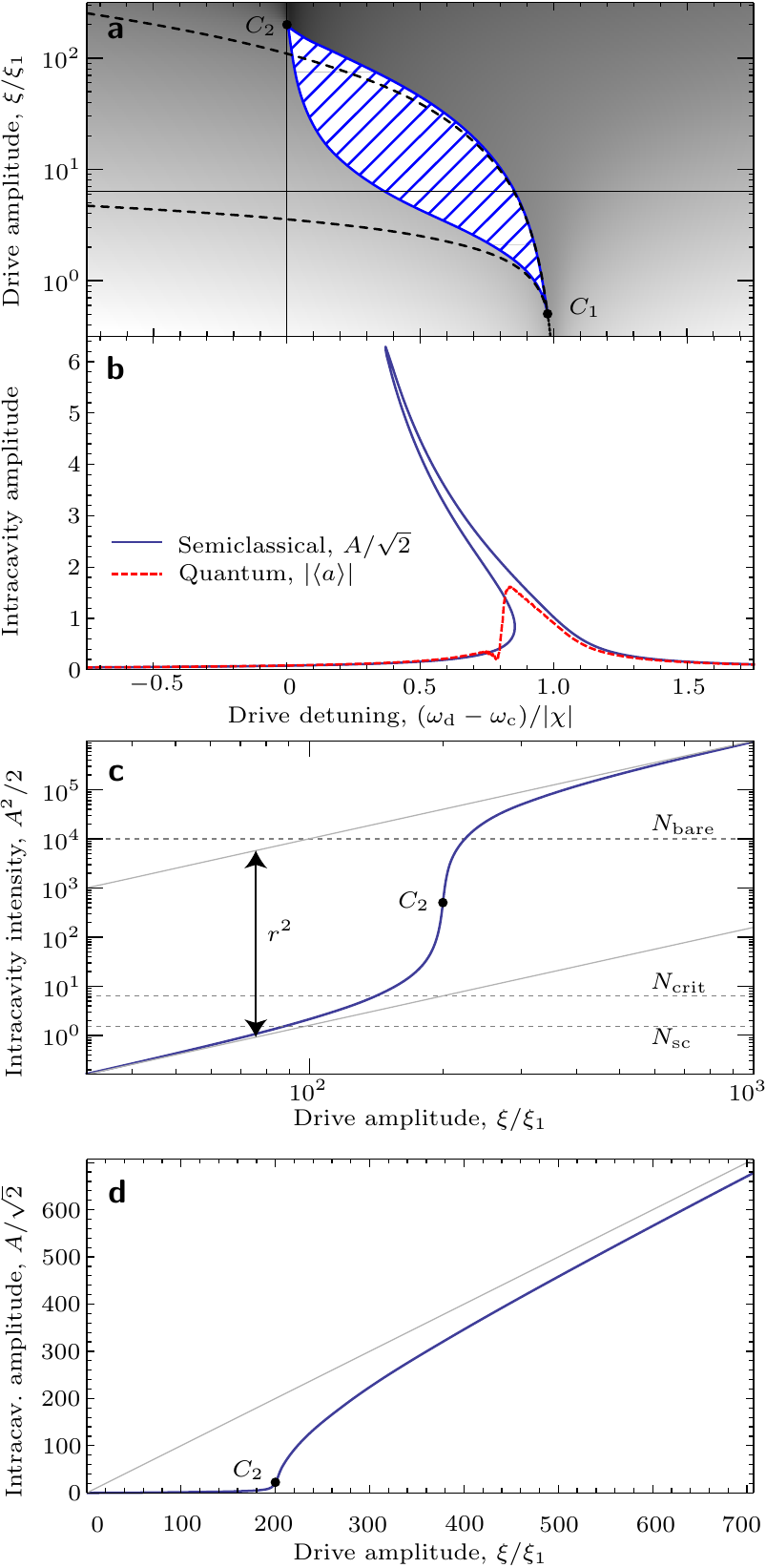}
\caption{(Color online). Solution to semiclassical equation~\eqref{eq:classic}, using the same parameters as Fig.~\ref{fig:latch000}b.
 (a)~Amplitude response as a function of drive frequency and amplitude, for same parameters as in Fig.~\ref{fig:latch000}b. The region of bifurcation is indicated by the shaded area, and has corners at the critical points $C_1$, $C_2$. The dashed lines indicate the boundaries of the bistable region for a Kerr oscillator (Duffing oscillator), constructed by making the power-series expansion of the Hamiltonian to second order in $N/\Ncrit$. The Kerr bistability region~\cite{manucharyan_microwave_2007} matches the JC region in the vicinity of $C_1$ but does not exhibit a second critical point.
 (b)~Cut through (a) for a drive of $6.3\xi_1$, showing the frequency dependence of the classical solutions (solid blue). For comparison, the response for from the full quantum simulation of Fig.~\ref{fig:latch000}b is also plotted (dashed red) for the same parameters.
 (c)~Cut through (a) for driving at the bare cavity frequency, showing the large gain available close to $C_2$ (the `step'). Faint lines indicate linear response.
 (d)~Same as (c), on a linear scale.\label{fig:densclass}}
\end{figure}

The solution of Eq.~\eqref{eq:classic} is plotted in Fig.~\ref{fig:densclass} for the same parameters as in Fig.~\ref{fig:latch000}. For weak driving the system response approaches the linear response of the dispersively shifted cavity. Above the lower critical amplitude $\xi_{C1}$ the frequency response bifurcates, and the JC oscillator enters a region of bistability. We denote by $C_1$ the point at which the bifurcation first appears. Dropping terms which are small according to the hierarchy of Eq.~\eqref{eq:regime}, this point occurs at
\be
    \xi_{C1}=\frac{(\delta\kappa)^{3/2}}{3^{3/4}g^{2}}, \quad \Omega_{C1}=\chi-\sqrt{3}\kappa/2 ,
\ee
writing the drive detuning as $\Omega=\omega_d-\omega_c$.
The dip in the heterodyne measurement of Fig.~\ref{fig:latch000} appears within the bifurcation region (Fig.~\ref{fig:densclass}a), indicating that this dip is the result of ensemble-averaging of the coherent heterodyne amplitude in the region of classical bistability. In Fig.~\ref{fig:densclass}b we see that the semiclassical and quantum simulation yield the same response outside the region of bistability. Within the region of bistability, quantum noise causes switching~\cite{dykman_fluctuations_1980} between the two semiclassical solutions, one dim and one bright, with almost opposite phases. Therefore, averaging over the amplitudes of an ensemble of independent realizations gives rise to a destructive interference effect at the frequency where the amplitudes weighted by the switching rates are similar. Other forms of single atom bistability are known: single atom absorptive bistability~\cite{savage} exists in the weak coupling regime in the good cavity limit, very different from the present discussion; more closely related is the single atom phase bistability of spontaneous dressed state polarization~\cite{alsing_spontaneous_1991,*kilin_single-atom_1991} which concerns the case where the atom and the cavity are in resonance $\delta=0$, unlike our situation where the detuning $\delta$ is the largest frequency scale apart from the cavity frequency.

As the drive increases, and unlike the Kerr oscillator, the frequency extent of the bistable region shrinks and eventually vanishes at the upper critical amplitude $\xi_{C2}=g/\sqrt{2}$. In the effective theory the upper critical point $C_2$ is located
very close to the bare cavity frequency. This indicates that for driving at the bare cavity frequency, there is no bistability, but rather a finite region (a `step') in the vicinity of the critical point (Fig.~\ref{fig:densclass}c), where the response becomes strongly sensitive to the drive amplitude. The size of the step can be shown to be a factor of $r=A_\text{bright}/A_\text{dim}=2 g^2/\kappa\delta$ in amplitude, and represents a very high gain ($\rmd A/\rmd \xi=\sqrt{2}g/\kappa^{3/2}\delta^{1/2}$) in the strong-dispersive regime. Above the step we see that the response approaches the linear response of the bare cavity as $N\simeq\Nbare$.

\begin{figure}
\includegraphics[width=0.8\columnwidth]{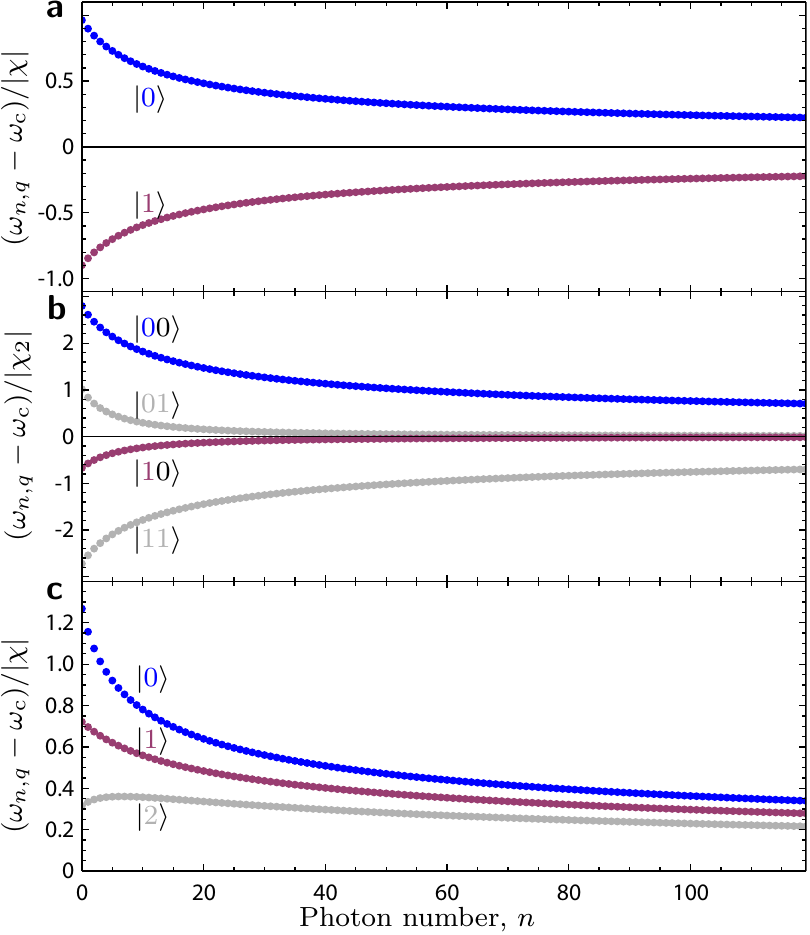}
\caption{(Color online) Symmetry breaking. State-dependent transition frequency versus excitation number:
(a)~for the JC model, parameters as in Figs.~\ref{fig:latch000} and~\ref{fig:densclass};
(b)~for the model extended to 2 qubits, $\delta_1=-1.0\,\text{GHz}$, $\delta_2=-2.0\,\text{GHz}$, $g_1=g_2=0.25\,\text{GHz}$;
(c)~for the model extended to one transmon qubit, tuned below the cavity, $E_C=0.2\,\text{GHz}$, $E_J=30\,\text{GHz}$, $g=0.29\,\text{GHz}$.
In all panels, the transition frequency asymptotically returns to the bare cavity frequency. In (a) the frequencies within the $\sigma_z=\pm1$ manifolds are (nearly) symmetric with respect to the bare cavity frequency. For (b), if the state of one (`spectator') qubit is held constant, then the frequencies are asymmetric with respect to flipping the other (`active') qubit. In (c), the symmetry is also broken due the existence of higher levels in the weakly anharmonic transmon.\label{fig:return}}
\end{figure}
From the semiclassical Eqs.~\eqref{eq:classic}, \eqref{eq:chi} it follows that for $A\gg 1$ the response of the system will have an approximate symmetry of reflection with respect to the bare cavity frequency $A(\Omega,\sigma_z=+1)\approx A(-\Omega,\sigma_z=-1)$. Therefore the response at the bare cavity frequency will be nearly independent of the state of the qubit.
In order to translate the high gain available at the step into a qubit readout, it is necessary to break the symmetry of the response of the system between the qubit ground and excited states, such that the upper critical power $\xi_{C2}$ will be qubit state dependent.
In the JC model the symmetry follows from the weak dependence of the decoupled Hamiltonian $\tilde{H}$ on the qubit state for high photon occupation. However, the experimentally-observed  state dependence may be explained by a symmetry breaking caused by the higher levels of the weakly anharmonic transmon, or by the presence of more than one qubit, see Fig.~\ref{fig:return}. When designing a readout scheme that employs such a diminishing anharmonicity, the contrast of the readout is a product of both the symmetry breaking and the characteristic nonlinear response of the system near the critical point $C_2$. Experiments~\cite{reedout_expt} were able to use this operating point to provide a scheme for qubit readout, which is attractive both because of the high fidelities achieved (approaching $90\%$, significantly better than is typical for linear dispersive readout in circuit QED~\cite{wallraff_approaching_2005,steffen_high_2010}) and because it does not require any auxiliary circuit elements in addition to the cavity and the qubit. During preparation of this manuscript we became aware of a theoretical modeling of the high-fidelity readout by Boissonneault \textit{et al.}~\cite{maxime_reedout}.

\begin{acknowledgments}
We acknowledge useful discussions with D.~I. Schuster,  R.~J. Schoelkopf 
and M.~H. Devoret. We thank M.~D. Reed and L.~DiCarlo for the data displayed in Fig.~\ref{fig:latch000}a. This work was supported by the NSF under Grants DMR-0603369 and DMR-0653377, LPS/NSA under ARO Contract W911NF-09-1-0514, and in part by the facilities and staff of the Yale University Faculty of Arts and Sciences High Performance Computing Center.
\end{acknowledgments}
\bibliography{levisordinary,levapsjour,reedout}
\end{document}

%% file: macros.tex
\newcommand*{\be}{\begin{equation}} 
\newcommand*{\ee}{\end{equation}}

\newcommand*{\omc}{\omega_{\text{c}}}
\newcommand*{\omq}{\omega_{\text{q}}}
\newcommand*{\omd}{\omega_{\text{d}}}
\newcommand*{\bigO}{\mathcal{O}}
\newcommand*{\Nbare}{N_\text{bare}}
\newcommand*{\Ncrit}{N_\text{crit}}
\newcommand*{\Nsc}{N_\text{sc}}

\newcommand*{\dg}{^{\dag}}

\newcommand*{\rmi}{\mathrm{i}}

\newcommand*{\rmd}{\mathrm{d}}